# Linear Regression Models in Epidemiology

Anatoly N. Varaksin, Vladimir G. Panov

*Institute of Industrial Ecology, the Urals Branch of the Russian Academy of Sciences,*
20, S. Kovalevskaya str., 620990 Ekaterinburg, Russia

varaksin@ecko.uran.ru
vpanov@ecko.uran.ru

## Abstract

The paper proposes to analyze epidemiological data using $Y = Y(x_1, x_2, ......, x_k)$ regression models which enable subject-matter (epidemiological) interpretation of such data whether with uncorrelated or correlated predictors. To this end, $Y = Y(x_1, x_2, ......, x_k)$ functions should include not only terms linear in predictors but also higher order ones (e.g. quadratic and cross terms). For epidemiological interpretation of a regression model, the suggestion is to construct functions of the type $Y = Y(x_i \mid \{x_j^0\})$ derived from the general regression function $Y = Y(x_1, x_2, ......, x_k)$ with the values of all predictor variables $x_j = x_j^0$ held fixed excepting one predictor, $x_i$. Unlike the conventional techniques based on linear-predictor models $Y = Y(x_1, x_2, ......, x_k)$ in which the *coefficient* for the $x_i$ variable is interpreted, our approach proposes to interpret the $Y = Y(x_i \mid \{x_j^0\})$ *function*, which is multivariate for any predictor $x_i$ being dependent on the values of all the other predictors $x_j$. It is such functions that can describe relationships between Y and $x_i$ of different forms in different predictor domains. The paper discusses differences in the interpretation of $Y = Y(x_i \mid \{x_j^0\})$ functions between cases involving correlated and uncorrelated predictor variables. The construction and analysis of regression models for epidemiological and environmental data are illustrated with examples.

**Keywords:** Epidemiology; Linear regression model; Cross terms; Conditional response functions; Model's coefficients; Model interpretation in epidemiology

**Introduction**

In many epidemiological and environmental studies relationships between the observed response $Y$ and the set of predictor variables $x_1, x_2, \ldots, x_k$ are established using multiple linear regression models of the following form (see, Rothman et al. 2008; Bonita et al. 2006)

$$Y(x_1, x_2, x_3, \ldots, x_k) = b_0 + b_1 x_1 + b_2 x_2 + b_3 x_3 + \ldots + b_k x_k \ , \qquad (1)$$

where $b_i$ ($i = 1, 2, \ldots, k$) are coefficients determined by the ordinary least squares method (Draper and Smith 1998, Rao and Toutenburg 1999). In statistics, the term "linear model" means that model coefficients $b_i$ enter equation (1) in a linear manner. Often the term "linear model" also implies that the predictors are linear (i.e. the predictor variables $x_i$ in equation (1) are not functions of any other variables).

Models (1) are used for solving three main types of problem:

1) A problem of predicting the observed values of a response Y by formula (1). If the resulting model is of high quality (i.e. with a high determination coefficient $R^2$), then the estimated values of Y are close to the observed ones, and formula (1) may be used for predicting observed values of Y by the known values of the set of predictor variables $x_i$.

2) A problem seeking to establish the relationship between *each* predictor variable $x_i$ and the response Y and provide a substantive interpretation of this relationship. It deals with the construction and interpretation of $Y = Y(x_i)$ functions derived from the basic equation (1) for each $x_i$ separately. Interpretation of the regression function means that the resulting functions $Y = Y(x_i)$ should be discussed in substantive epidemiological terms (how Y changes with a given $x_i$). Such problems are popular with epidemiologists, environmentalists and other scientists where the response of a system (for instance, morbidity) needs to be explained with the help of a set of predictor variables (for instance, a set of data on environmental pollution by various toxic agents $x_i$). In this case, $Y = Y(x_i)$ functions should be "translated" into the language of medicine, ecology, etc. For instance, one is required to demonstrate how morbidity changes with each specific environmental pollutant. Moreover, it is essential that the obtained regression models and $Y = Y(x_i)$ functions should be consistent with specific (epidemiological, environmental) premises (for instance, that morbidity should increase with environmental pollution).

3) A problem of determining how predictors (factors) act jointly in a combination. A problem like this occurs more frequently in designed biological (toxicological) experiments, but it may also arise in a "non-designed" (observation) study. Where two factors are simultaneously impacting on a system, the problem is specified as follows: is the action of two factors additive or not (if not, then what kind of action is it?). It will be recalled that an action is considered to be additive when the effect of two factors acting together is equal to the sum of their single-factor actions in the same doses. From this viewpoint, Type (1) model linear in predictors (factors) describes only one type of combined action produced by these factors, which is additive. The action of two factors is non-additive when the action of one of the factors is different at different levels of the other factor. To be able to describe non-additive effects in regression terms, a model should contain terms which are non-linear in predictors (more precisely, cross terms of the $x_i \cdot x_j$ type).

For solving the first problem, it would be sufficient to construct multiple linear regression (MLR) of high quality, i.e. a model with a high value of the determination coefficient $R^2$ or a low value of the residual sum of squares (RSS). If this is achieved, the problem is considered to be solved (Draper and Smith 1998); otherwise one would have to resort to other methods for constructing a predictive model (for instance, more complex non-linear variants of regression or neural networks).

Epidemiology mainly deals with problems of the second type. It may seem that such problems should not present any difficulty for resolving them by regression methods. Indeed, according to (1) the relationship between $Y$ and a specific predictor $x_i$, with the values of the other predictors $x_j^0$ held fixed, is given by

$$Y(x_i \mid x_j^0) = B_0 + b_i x_i, \qquad (2)$$

where $B_0$ is a constant determined by the free term $b_0$ of formula (1) and by the product $b_j x_j^0$, the values of the other predictors $x_j^0$ held fixed.

The common epidemiological interpretation of the relationships (1) and (2) is as follows. Let, for instance, $Y$ be a population morbidity caused by environmental pollution with certain toxic agents in concentrations $x_i$. It is necessary to answer the question which of the pollutants influence the morbidity $Y$ and how. From equations (1)-(2) it follows that a unit increase in the concentration of the pollutant $x_i$ increases the morbidity $Y$ by a value equal to $b_i$:

$$Y(x_i + 1) - Y(x_i) = \Delta Y(\Delta x_i = 1) = b_i . \qquad (3)$$

Thus, the relationships (2)-(3) suggest that a change in $Y$ with a change in $x_i$ depends only on the coefficient $b_i$ and does not depend on the values of the other predictors $x_j$ (Rothman et al. 2008). This interpretation of model (1) obviously implies that all predictors can vary independently of one another. In reality, this is only possible in a case of uncorrelated predictor variables. Where there is correlation between such variables, a change in one of them changes all the others.

As for the third problem, it is imperative to include terms which are non-linear in predictors (in particular, cross terms containing products of the type $x_i \cdot x_j$) into the MLR model. The testing of cross terms may reveal their statistical significance (in this case, the actions of $x_i$ and $x_j$ will be non-additive) or insignificance (the joint action of the factors will be additive).

It is important to note that the construction of various regression models, whether linear or non-linear in predictors and coefficients, has been well-developed and described in numerous monographs and papers on applied regression analysis. Suffice it to mention the latest edition of the fundamental monograph by Draper and Smith (1998) and the comprehensive monograph by Rothman et al. (2008). These publications contain a detailed description of regression equation construction techniques, statistical hypothesis testing, and other largely statistical questions. However, they do not give sufficient attention to substantive interpretation of regression models.

As a mathematical discipline, regression analysis deals with both general issues in regression modeling and the properties of specific regression models. The general issues are concerned with the properties of regression models which do not depend on a specific analytical form of the regression function. On the contrary, discussions of specific regression models (regression functions) focus on their special features (Rothman et al. 2008). In this paper, we only deal with some special regression models which are frequently used in epidemiology, environmental studies, and response surface theory  (see, e.g. Myers et al. 2016 and references therein, and Castelleti et al. 2010).

This study seeks:

- to show which regression models are more appropriate for environmental health (epidemiological) studies in the presence or absence of correlations between predictor variables;

- to demonstrate examples of substantive (epidemiological) interpretation of regression models using examples from epidemiological (biomedical) research.

## 1. A regression model with correlated predictor variables.

The common substantive interpretation of Type (1) predictor-linear regression models described by the relationships (2)-(3) above is widely used in epidemiology, environmental health (and other) studies (see, e.g., Bonita et al. 2006). The authors of this monograph comment that the coefficient $b_i$ in formula (3) describes a change in Y with a change in $x_i$ *adjusted for all other terms in the model* $x_j$ (highlighted by us). The monograph of Bonita et al. (2006) does not explain the meaning of the term "adjusted" in the context of *regression*, i.e. no strict mathematical definition is given for regression adjustment. It is clear that the discussion relates to a model with correlated predictor variables; otherwise coefficients $b_i$ would not depend on the presence of other predictor variables and adjustment would make no sense.

Let us try to understand what happens to the coefficients of a regression model when new predictor variables are included into it with the result that these coefficients undergo a change. Let there be two predictors $x_1$ and $x_2$, for which we can construct two simple linear regression (SLR) equations:

$$Y = a_{0,1} + a_1 x_1 \,, \qquad Y = a_{0,2} + a_2 x_2 \,. \tag{4}$$

If an MLR equation is constructed with the same predictor variables and the same outcome Y

$$Y = b_0 + b_1 x_1 + b_2 x_2 \tag{5}$$

then in the presence of correlations between $x_1$ and $x_2$ the coefficients $b_1$ and $b_2$ of the MLR model (5) appear to be different from the coefficients $a_1$ and $a_2$ of the simple regression models (4). It is this change of regression coefficients in passing from SLR to MLR that seems to be taken for adjustment.

It is to be emphasized once again that such regression adjustment is not defined either mathematically or epidemiologically. Mathematically, the transition from $a_1$ and $a_2$ to $b_1$ and $b_2$ occurs according to the following rules. Let us introduce notation for the regression equations relating predictors $x_1$ and $x_2$ to each other:

$$x_1 = c_{0,1} + c_{12} x_2 \,, \tag{6}$$

$$x_2 = c_{0,2} + c_{21} x_1 \,. \tag{7}$$

In these notations, the relationship between the one-factor and two-factor coefficients is given by Panov and Varaksin (2015):

$$b_1 = \frac{a_1 - a_2 c_{21}}{1 - r^2(x_1, x_2)} \,, \tag{8}$$

$$b_2 = \frac{a_2 - a_1 c_{12}}{1 - r^2(x_1, x_2)} \,, \tag{9}$$

where $r(x_1, x_2)$ is a Pearson correlation coefficient between $x_1$ and $x_2$. For uncorrelated predictors, the coefficients $c_{21}$, $c_{12}$ and $r(x_1, x_2)$ are equal to zero, and consequently, $b_1 = a_1$, $b_2 = a_2$ and no adjustment takes place.

Formulas (8) and (9) show how, for example, the coefficient $a_1$ of the regression model containing one predictor $x_1$ changes when a predictor $x_2$ is added. In this formula, there is no indication why the transformation of the coefficient $a_1$ into $b_1$ should be considered as an act of adjustment (because "adjustment" is not clearly defined).

Formulas (6)–(9) explain how the coefficients $a_1$ and $a_2$ turn into $b_1$ and $b_2$. There is no epidemiological interpretation of the coefficients $b_i$ at all. Some mathematical sense of the coefficients $b_i$ in MLR (1) gives the following result. Following the ideas of Mosteller and Tukey (1977, Ch. 13), we showed (Panov and Varaksin 2015) that the coefficient $b_1$ in MLR (5) is equal to the coefficient $a_1^*$ in the simple regression equation

$$Y = a_0^* + a_1^* x_1^*,\tag{10}$$

where the new variable

$$x_1^* = (x_1 - c_{12} x_2),\tag{11}$$

while the coefficient $c_{12}$ describes a linear regression relationship between the predictors $x_1$ and $x_2$ as in formula (6). The substantive meaning of the coefficient $b_1 = a_1^*$ for the predictor $x_1$ in equation (5) is now clear: $b_1$ is a coefficient for the relationship between $Y$ and the predictor $x_1$ from which the linear part of its relationship with the predictor $x_2$ has been deleted, i.e. $x_1^*$ is the $x_1$ predictor "stripped" of the effect of $x_2$ ($x_1^*$ and $x_2$ are not correlated). Where many predictors are present, the coefficient $b_1$ for the predictor $x_1$ will be equal to the coefficient $a_1^*$ in equation (10), where

$$x_1^* = x_1 - c_{12} x_2 - c_{13} x_3 - .... - c_{1k} x_k\tag{12}$$

i.e. $x_1^*$ is the $x_1$ predictor from which its linear relationships with the other predictors have been deleted. We can interpret similarly any predictor-linear coefficient $b_i$ in equation (1).

As well as in the approach based on formulas (8) - (9), in this approach we do not see any epidemiological sense. Moreover, if the epidemiological interpretation of model (1) is used in the form of the relationships (2) and (3), then the response Y of a complex system to an impact $x_i$ is described by *one* relationship of the type $Y = b_0 + b_i x_i$ over the *entire* range of changes in $x_i$ for *any* values of the other predictors $x_j$ (irrespective of the presence or absence of correlations between the predictors). Such an oversimplified description of the system is unlikely to be true in most cases (illustration see below, example 3). We may also refer to the area of toxicology dealing with combined effects of toxicants. In the majority of cases, each of the toxicants acts differently in the different dose ranges of the other toxicants, and the impact of any toxicant $x_1$ is not described by the same formula $Y = b_0 + b_1 x_1$ over the entire range of changes in the other toxicant $x_2$ (Calabrese 1991; Minigalieva et al. 2017); these differences cannot be described in principle by a regression model linear in predictors.

Nonetheless, the above-described approach to substantive (epidemiological) interpretation of Type (1) regression equations by relations (2)-(3) persists among healthcare professionals and environmentalists. Such an (incorrect) substantive interpretation of equation (1) can be found, for instance, in the monograph of Bonita et al. (2006), published by the World Health Organization.

That is not to say that all epidemiologists agree with the substantive interpretation of equation (1) by relationships (2)-(3). A different approach is employed in the article of Abbott and Carroll (1984). For estimating the effect $\Delta Y(x_i)$ produced by some predictor (e.g., the first predictor $x_1$), these authors replaced $\Delta Y(x_1) = b_1$ with the relationship

$$\Delta Y(\Delta x_1 = 1) = b_1 + b_2 c_{2,1} + b_3 c_{3,1} + \cdots , \qquad (13)$$

where $c_{i,j}$ represents the coefficients of simple regression equations of the type (6) - (7) which link the predictors $x_1$ and $x_j$: $x_j = c_{0,j} + c_{j,1} x_1$. The substantive meaning of formula (13), according to Abbott and Carroll (1984), is as follows. If the predictor $x_1$ were not correlated with the other $x_j$, then the effect $\Delta Y(\Delta x_1 = 1)$ due to a unit change in $x_1$ would be equal to $b_1$. As a consequence of the correlations between $x_1$ and $x_j$, a unit change in $x_1$ induces a change in the «means» of the other predictors $x_j$ by the amount $c_{j,1}$. According to (1), The change in $x_j$ by the amount $c_{j,1}$, leads, in turn, to a change in Y by the amount $b_j c_{j,1}$, which is what formula (13) reflects.

These reasonings seem to be logical; however, as was shown in Panov et al. (2008) and Panov and Varaksin (2010), for a regression which is linear in the predictors, summation in (13) gives

$$\Delta Y(\Delta x_1 = 1) = a_1 , \qquad (14)$$

where $a_1$ is a coefficient of the simple regression equation $Y = a_0 + a_1 x_1$ relating Y to one (first) predictor only. This means that the effect of the predictor $x_1$ adjusted in a multifactorial model for the correlations between the predictors proves to be equal to the non-adjusted effect in the simple linear regression model. Thus, this attempt to adjust the effect of the predictor $x_i$ allowing for its correlations with the other predictors $x_j$ should be recognized as unsuccessful, and we have to find other ways to construct and interpret Type (1) equations if there are correlations between the predictors $x_i$.

## 2. Models with the cross terms

Cross-term of the type ($x_i \cdot x_j$), in which predictors are multiplied, is one of the first terms where the form of the regression model becomes more complex due to higher terms than ones linear in $x_i$. The inclusion of other terms non-linear in predictors into the model (for instance, terms quadratic in $x_i$) will be considered below. Let us start with a simple case of regression with two predictors $x_1$ and $x_2$:

$$Y = b_0 + b_1 x_1 + b_2 x_2 + b_{12} x_1 x_2. \qquad (15)$$

Compared with equation (5), it has a term containing the product of the predictors $x_1$ and $x_2$. Equations with cross-terms are well-known in the regression theory (see, e.g., Draper and Smith, Ch. 7; Rothman et al. 2008, Ch. 20). In this paper, we are interested in the question: how could equation of the type (15) be used for substantive interpretation of the impact of each of the factors $x_1$ and $x_2$ on some characteristic of the system Y in the presence of correlations between $x_1$ and $x_2$?

As was stated in the Introduction, it is necessary to construct and interpret one-factor functions $Y = Y(x_i)$ derived from (15), the values of the other predictor held fixed. For obtaining $Y = Y(x_i)$, let us fix in (15) the value of the predictor x2 on some level $x_2 = x_2^0$ (this procedure is similar to the construction of equation (2)). The result is a relationship between Y and the first predictor $x_1$ given by:

$$Y = Y(x_1 \mid x_2^0) = (b_0 + b_2 x_2^0) + (b_1 + b_{12} x_2^0) \cdot x_1. \qquad (16)$$

As well as in the relationship (2), the relationship between Y and the predictor $x_1$ proves to be linear in $x_1$. However, the coefficient for $x_1$ depends on the fixed value of $x_2^0$ for which we estimate the relationship between Y and $x_1$: The latter proves to be different for different values of $x_2^0$ (Rothman et al. 2008, Ch. 20).

Let us write by analogy with (3) a change in the response Y with a unit change in the first predictor $x_1$ for the fixed values of the predictor $x_2 = x_2^0$. From (15), it follows that

$$Y(x_1 + 1) - Y(x_1) = \Delta Y(\Delta x_1 = 1) = b_1 + b_{12} x_2^0 \; , \qquad (17)$$

i.e. this change, too, depends on $x_2^0$ and the value of $\Delta Y(\Delta x_1 = 1)$ proves to be different for different values of $x_2^0$ (Rothman et al. 2008). Moreover, in the case of correlated predictors, the inclusion into (15) of a cross-term changes the values of the coefficients $b_1$ and $b_2$ for linear terms: The coefficients $b_1$ and $b_2$ in equation (15) prove to be other than in equation (5).

Thus, the inclusion of the cross-term $b_{12} x_1 x_2$ into the regression equation changes the *epidemiological interpretation* of the equation regression due to both the inclusion of a new addend (cross-term) and a change in the coefficients $b_1$ and $b_2$ (if the term $x_1 x_2$ correlates with $x_1$ and/or $x_2$). Moreover, epidemiological interpretation of the effect of the predictor $x_i$ on Y is based on the formulas of the type (16) rather than on separate coefficients $b_i$, as it is assumed in Bonita et al. (2006). This also illustrates by Example 1 (see below).

Could the inclusion of term $b_{12} x_1 x_2$ present a method specifically for taking into account correlations between predictors $x_1$ and $x_2$? Explicitly, no such consideration is obvious, but it is done indirectly! Indeed, an explicit indicator of a correlation between $x_1$ and $x_2$ is the correlation coefficient $r(x_1, x_2)$, but in the formulas (15)-(17) there is no such correlation coefficient. At the same time, formulas (15)-(17) do allow for the correlations between $x_1$ and $x_2$ indirectly as follows: in the absence of a cross-term, the response function $Y = Y(x_1)$ depends only on $x_1$ (an *unconditional* relationship between Y and x₁), while in the presence of $b_{12} x_1 x_2$ it turns into a *conditional* relationship $Y = Y(x_1 \mid x_2^0)$ where the relationship between Y and $x_1$ depends on $x_2^0$; so we have a possibility to allow for a change in x₂ with a change in x₁ when estimating $\Delta Y$ (17). Inclusion of a cross-term into a regression model renders it more flexible enabling one to better allow for the relationship between factors at hand. The fact that this may involve a cardinal change in the substantive interpretation of the model is a consequence of not only (and not always) correlation predictors but also of a better description of experimental data by the regression model with a cross-term.

## 3. A regression model with non-correlated predictors.

If interpretation of a regression model with correlated predictors causes certain difficulties (because of correlations between the predictors), it may be expected that interpretation of Type (1) regression equations with non-correlated predictors is unlikely to cause any trouble. Indeed, in the absence of any correlations between the predictors, the latter may change independently of each other. In this case, the inclusion into the model of new predictors $x_j$ would not change the coefficients $b_i$ of the already existing predictors. Therefore a unit change in some predictor $x_i$ leads to a change in Y by the amount $b_i$ irrespective of the other predictors $x_j$.

It may seem that the question of substantive interpretation of Type (1) regression models with non-correlated predictors has been resolved uniquely and correctly from the standpoint of both mathematics and subject area (epidemiology). However, the actual situation is not so simple. One example is the field of so-called designed experiments in which the design does not provide for any correlation between the predictors but a model which is linear in predictors proves to be explicitly insufficient for describing adequately the experimental data. This situation is corrected by including $(x_i \cdot x_j)$ cross-terms into the regression. Provided the cross-term is sufficiently statistically significant, substantive interpretation of a model with this cross-term differs essentially compared with a model linear in predictors. We described this variant in the publications (Varaksin et al. 2014; 2018; Panov and Varaksin 2016). One of the examples is given below (example 2).

Interestingly, following the inclusion of cross-terms into a regression model, its substantive interpretation will be identical for both correlated and non-correlated predictors. Differences between these cases will only occur in the allowable range of predictor variation which fits the experimental data. In the case of non-correlated predictors, any combinations of predictor values are possible (the conditional probability $W(x_2 \mid x_1)$ does not depend on $x_1$). In the presence of correlations, some combinations of $x_1$ and $x_2$ values are not observed in experiment (or occur less often than others). For instance, in the case of a positive correlation between $x_1$ and $x_2$, low values of $x_1$ will be highly unlikely for large $x_2$ and vice versa (see Fig.1). As the correlations between the predictors get stronger, the joint domain of model predictors $\{x_j\}$ narrows down (the confidence ellipse of the observed predictor values becomes increasingly elongated). At some value of correlation coefficient, it may become so thin (narrow) that the sample $\{Y, x_j\}$ becomes non-representative (does not relate to all possible values of $Y$ and $\{x_j\}$). The function $Y = Y(x_1, x_2, \ldots, x_k)$ defined in this small (thin) joint domain $\{x_j\}$ will become «unstable», and paradoxical results of regression interpretation will be inevitable (Ehrenberg 1975; see also example 1).

## 4. Models with cross and quadratic terms

Let us add quadratic terms to model (15). The inclusion of these terms presents another possibility for improving model quality. We obtain a complete quadratic function in $x_1$ and $x_2$:

$$Y = b_0 + b_1 x_1 + b_2 x_2 + b_{12} x_1 x_2 + b_{11} x_1^2 + b_{22} x_2^2 \quad . \tag{18}$$

If in (18) we hold the value of one of the predictors $x_2 = x_2^0$ fixed (by analogy with the construction of equation (2)), the result is a relationship between $Y$ and predictor $x_1$ given by:

$$Y = Y(x_1 \mid x_2^0) = (b_0 + b_2 x_2^0 + b_{22} (\mathrm{x}_2^0)^2) + (b_1 + b_{12} x_2^0) \cdot x_1 + b_{11} \cdot x_1^2 . \tag{19}$$

Now the relationship between $Y$ and predictor $x_1$ becomes quadratic, and the coefficient of the term linear in $x_1$ depends on the fixed value of $x_2^0$ for which the relationship between $Y$ and $x_1$ is estimated.

Let us write, by analogy with (3), a change in the response Y with a unit change in the first predictor $x_1$ for the fixed values of the predictor $x_2 = x_2^0$; from (19), it follows that

$$Y(x_1 + 1) - Y(x_1) = \Delta Y(\Delta x_1 = 1) = b_1 + b_{12} x_2^0 + b_{11} (2x_1 + 1) \tag{20}$$

Now the value of $\Delta Y(\Delta x_1 = 1)$ depends not only on $x_2^0$ (as it was in the model without quadratic terms) but on $x_1$ as well. An example illustrating the change in the substantive interpretation of the relationship between the response and the predictors with quadratic terms included into the model was described by us in Varaksin et al. (2018).

## 5. General propositions for regression models with arbitrary predictors

For epidemiological studies, it is proposed to use regression models $Y = Y(x_1, x_2, \ldots, x_k)$ that contain not only linear in explanatory variables, but also terms of higher order (cross, quadratic and other terms). The epidemiological interpretation of such models is based on an analysis of one-factor functions $Y = Y(x_i \mid \{x_j^0\})$ derived from $Y = Y(x_1, x_2, \ldots, \mathrm{x}_k)$ for each predictor $x_i$. If the regression model $Y = Y(x_1, x_2, \ldots, \mathrm{x}_k)$ describes the experiment well, then the values of the response Y calculated by the regression model are found to be close to experimental values. In this case, the function $Y = Y(x_i \mid \{x_j^0\})$ should adequately describe the relationship between Y and a given $x_i$. Thus, the construction of MLR of high quality (high $R^2$, low RSS) is an essential stage required for correct interpretation of regression models.

If the researcher has variables $x_i$ ($i = 1, 2, \ldots, k$) which may serve as predictors in the MLR, the general form of the MLR model may be given by

$$Y = b_0 + \sum_i b_i x_i + \sum_i \sum_{j>i} b_{i,j} x_i x_j + \sum_i b_{ii} x_i^2 + \sum_i \sum_{j>i} \sum_{l>j} b_{i,j,l} x_i x_j x_l + \sum_i b_{iii} x_i^3 + \sum_i \sum_{j \neq i} x_i^2 x_j + \dots \quad (21)$$

where the first sum represents terms which are linear in predictors, the second sum represents paired cross-terms, the third sum terms which are quadratic in predictors, etc. (terms to the third order inclusive are written up explicitly).

The need to include into the regression model terms which are non-linear in predictors (particularly cross-terms) in case of their significance has been known since long (see, e.g., Draper and Smith 1998; Rothman et al. 2006). In our paper, it is emphasized that the inclusion of non-linear terms into MLR not only improves the quality of the model (increase in $R^2$ and decrease in RSS) but also modifies the interpretation of the model (radically in some cases). It is the correct substantive interpretation on the basis of the functions $Y = Y(x_i \mid \{x_j^0\})$ is the goal of statistical analysis of epidemiological data. To this end, it is essential to have an adequate (in relation to experimental data) regression model $Y = Y(x_1, x_2, \dots, x_k)$. It is the lack of this model and/or incorrect interpretation of oversimplified models (linear in predictors) is a common mistake in regression analysis of epidemiological data. Interpretation of such simplified models constructed without allowing for their substantive meaning results in epidemiologically "paradoxical" conclusions (Ehrenberg 1975). The most characteristic and frequent example of a paradoxical conclusion is a reduction in the morbidity with increased environmental pollution (see, for example, Panov and Varaksin 2015). In so doing, a paradoxical finding is, as a rule, a consequence of predictor correlations in the simplified regression model (the model lacks cross and other high order terms). A conclusion may also appear paradoxical where an attempt is made to interpret the effect of a secondary (statistically insignificant) predictor on the response.

Concerning practical applications of type (21) MLR, the following recommendations may be proposed.

1) If a primary analysis of experimental data does not explicitly point to the presence of higher-order non-linearities in the dependence of Y on predictors $x_i$, form (21) may be limited to the first three terms (quadratic polynomial). This polynomial (21) should include statistically significant cross-terms of substantive (main) predictors and products of these predictors with the rest. The *hierarchy principle* (Bishop et al. 1975) recommends include in the model linear terms of predictors which are presented in cross-terms.

In some particular cases, regression models may have higher-order terms of the type ($x_i \cdot x_j \cdot x_l$), with the quadratic and cubic terms (types $x_i^2$ and $x_i^3$) in them being completely insignificant. We described such a situation in (Katsnelson et al. 2015). In this case, one should be guided by the substantive significance of the explanatory independent variables.

If a constructed MLR $Y = Y(x_1, x_2, \ldots\ldots, x_k)$ is of high quality, it may be used for constructing conditional dependencies $Y = Y(x_i \mid \{x_j^0\})$, a correct epidemiological interpretation of which may be sufficient in many cases.

2) For constructing MLR, there is an empirical recommendation ($k < n/10$) to limit the number of predictors $k$ for a given number of observations $n$ (Rothman et al. 2008). Typically, researchers try to find a formula that would provide a maximum $R^2$ with minimum predictors. To this end, a lot of algorithms have been developed and described (Rothman et al. 2008). Provided the limitation $k \ll n$ is observed, the violation of the rule $k < n/10$ should not tell on the epidemiological interpretation of the model. A greater risk than the violation of the rule $k < n/10$ comes from correlations between predictors. If a certain level of coefficients of correlation between $x_i$ is exceeded, then one would not be able to save the regression model even by including cross-terms in it with the result that the model would be producing paradoxical results.

3) If the number of predictors of interest in a given problem is not large ($k$ from 3 to 5), the epidemiologist may include into the MLR model (21) linear, quadratic and paired cross-terms of all predictors irrespective of their statistical significance, provided, however, their epidemiological significance is taken into account. The main thing is to ensure that the model is of high quality and, thus, describes adequately the experimental data. Then the functions $Y = Y(x_i \mid \{x_j^0\})$ will correctly describe the relationships between the response Y and all predictors $x_i$.

If confined to the quadratic form, the function $Y = Y(x_1 \mid \{x_j^0\})$, for example for the predictor x₁, will be given by:

$$Y = Y(x_1 \mid \{x_j^0\}) = T_0 + T_1 x_1 + T_{11} x_1^2, \qquad (22)$$

where (using the notations of equation (21))

$$T_0 = b_0 + \sum_{i=2}^{k} (b_i x_i^0 + b_{ii}(x_i^0)^2) + \sum_{i=2, j>i}^{k} b_{ij} x_i^0 x_j^0, \qquad T_1 = (b_1 + \sum_{i=2}^{k} b_{1,i} x_i^0), \qquad T_{11} = b_{11}. \qquad (23)$$

Thus, the coefficient $T_1$ in (22) describing a linear relationship between Y and x₁ is determined by the coefficient $b_1$, coefficients $b_{1,i}$ and the values of $x_i^0$ which, if held fixed, give the value of $T_1$. There may be a lot of various combinations of fixed values of $x_i^0$, and each combination $x_i^0$ would have its value of $T_1$. In practical terms, one may select the most «interesting» combinations of $x_i^0$ values for analysis, for instance, a set of predictor means. In choosing combinations of $x_i^0$ values, it is important to allow for possible correlations between the predictors, as these may render some combinations of { $x_i^0$ } more or less probable. Of special interest for substantive interpretation are cases where a change in $x_j^*$ changes dramatically the form of the function $Y(x_1 \mid x_2^0, .... x_k^0)$ .- заменить звезды на нули.

4) If the initial number of predictors is large and the form of (21) becomes two "cumbersome", one can exclude from the quadratic form (21):

- statistically insignificant cross-terms which do not contain important predictors;

- statistically insignificant linear terms which do not appear in the cross-terms remaining in the model.

The exclusion of insignificant predictors may be performed with the help of the known procedure of backward stepwise regression. Note that after the exclusion of insignificant predictors the regression coefficients should be recalculated.

Note, however, that the inclusion or exclusion of variables should be performed not only on the basis of their statistical significance but also allowing for the role these variables play in the substantive explanation of a phenomenon studied. From the formal viewpoint of a mathematical model, all variables play a similar role. However, when we allow for the epidemiological contents of the variables and their role in the description of the system, they will form a certain hierarchy, which should be taken into account in the first instance in applied research.

## Conclusion

In 2006, the World Health Organization published a new edition of the monograph "Basic Epidemiology" by Bonita et al. (2006). Thus, the methodology of regression model construction and interpretation described in it may be regarded as recommended for use by epidemiologists across the globe.

The section 5 of this monograph (Bonita et al. 2006, pp. 76-77) devoted to regression analysis states with reference to type (1) relationships: "If the independent variable is a continuous variable … , then the *interpretation* of $b_i$ is straightforward and represents the incremental change in the dependent variable Y … , associated with a unit change in $x_i$ , *adjusted for all other terms in the model.*" In the case of *correlated* predictors, this statement is incorrect (see above sections 1-3 and the monograph Mosteller and Tukey (1977). In the case of *non-correlated* predictors, a regression model linear in predictors may prove to be insufficient for a "correct" description of relationships between Y and $x_i$ . Incidentally, the section 5, concerning regression analysis in the monograph Bonita et al. (2006) does not discuss the problem of correlated predictors at all.

As for the term "adjusted", the monograph Bonita et al. (2006) does not explain its meaning in the context of *regression*, i.e. no clear mathematical definition is given for the procedure of regression adjustment. The term " adjustment" in epidemiology is often associated with the term "standardization" in comparative studies, which implies "matching" of *two groups* of observations by the values of covariates using data stratification and "standard population" (Bonita et al. 2006; Ahmad et al. 2001). The procedure of standardization as a method of "matching" is well defined mathematically and is clear to the epidemiologists. In regression models, we can see no signs of "matching" and use of "standard population" in passing from SLR to MLR.

The methodology of constructing MLR of the type (21) with the inclusion of cross-terms allows one to interpret regression models epidemiologically without resorting to terms such as "adjustment". Epidemiological interpretation of models proves to be similar for regression with both non-correlated and correlated predictors. For moderate correlations between predictors, a model with cross-terms provides quite a reasonable epidemiological interpretation of the effect of each predictor on the response Y. As correlations between the predictors grow stronger, the probability of "paradoxical" inferences increases but does not make them inevitable.

Additional adjustment procedures are not needed only in the above-described regression situations (non-comparative studies) where all predictors in a regression model are *a priori* equal and it is necessary to describe the effect of each of them on the response Y. Regression adjustment procedures of the standardization type are mandatory in comparative studies. For instance, where it is necessary to determine the effect size of some exposure when the populations prove to be different for different values of exposure judging by concomitant variables (Rothman et al. 2008).

It should be emphasized once again that the epidemiological interpretation of a regression model should be based on an analysis of $Y = Y(x_i \mid \{x_j^0\})$ *functions* rather than on an analysis of individual regression *coefficients* $b_i$. A similar opinion was put forward by Mosteller and Tukey, 1977 (Chapter 13). The very possibility of analyzing $Y = Y(x_i \mid \{x_j^0\})$ *functions* rather than $b_i$ *coefficients* occurs in a multiple regression mode only after the inclusion of cross-terms into it. In the absence of cross-term terms, the contribution of each predictor to the response Y is described by one type relation over the entire range of variation for any values of the other predictors. Such an oversimplified description of the system can hardly be true in most cases.

# Examples

**Example 1.** A regression model with two correlated predictors.

Theme: Association between morbidity and air pollution in Saint-Petersburg (Russia).

There are data on the general morbidity of the population in 19 districts of Saint-Petersburg depending on the concentration of 12 air pollutants (CO, $NO_2$, $SO_2$, Pb, etc.). Experimental data are taken from (Scherbo (2002)). The problems to be resolved are:
- to establish a relationship between morbidity Y and pollutants concentration X;
- to determine the effect of each of the pollutants on morbidity and give an epidemiological interpretation of the regression equation.

The handling of the first problem commonly starts with a calculation of the Pearson coefficients of correlation between morbidity Y and predictors $x_i$. It is necessary to construct a simple regression equation linking the morbidity Y with each of the pollutants $x_i$ separately. Estimates of the correlations between the pollutants could also be useful.

Table 1. Pearson coefficients of correlation between general morbidity Y and some predictors $x_i$, as well as between the predictors $x_i$.

| Indices | Y | CO | NO2 | PM | Formaldehyde | Pb | Xylol | SO2 |
|---|---|---|---|---|---|---|---|---|
| Y (morbidity) | | 0.578 | 0.544 | 0.409 | 0.403 | 0.376 | 0.259 | 0.121 |
| CO (carbon monoxide) | 0.010 | | 0.754 | 0.492 | 0.588 | 0.302 | 0.682 | 0.729 |
| NO2 (nitrogen dioxide) | 0.016 | <0.001 | | 0.484 | 0.582 | 0.374 | 0.459 | 0.552 |
| PM (particulate matter) | 0.082 | 0.032 | 0.036 | | 0.739 | 0.304 | 0.087 | 0.233 |
| Formaldehyde ($CH_2O$) | 0.087 | 0.008 | 0.009 | <0.001 | | 0.343 | -0.037 | 0.263 |
| Pb (lead) | 0.112 | 0.210 | 0.115 | 0.206 | 0.151 | | -0.015 | 0.286 |
| Xylol ($C_8H_{10}$) | 0.285 | 0.001 | 0.048 | 0.724 | 0.882 | 0.951 | | 0.801 |
| SO2 (sulfur dioxide) | 0.622 | <0.001 | 0.014 | 0.337 | 0.277 | 0.235 | <0.001 | |

In the table above the diagonals are correlation coefficients and those under the diagonals are corresponding p-values. The toxicants in the table are sorted in the decreasing order of correlation coefficients between Y and $x_i$. The strongest and statistically most significant (p<0.05) correlations are observed between the morbidity and the concentrations of carbon monoxide CO and nitrogen dioxide $NO_2$; the correlations are weakest between the morbidity Y and the concentrations of $SO_2$: $r(Y, SO_2) = +0,121$ ($p = 0,622$).

The strongest correlations between the pollutants are for the pairs (CO; $NO_2$) and (CO; $SO_2$). Relatively weak correlations with all the other predictors are demonstrated by the toxicant Pb.

Simple regression equations needed further on are given by (the numbering of the formulas is continued from the main body of the paper)

$$Y = 603 + 579 \; CO, \tag{24}$$

$$Y = 919 + 52{,}5 \; SO_2. \tag{25}$$

where Y is morbidity (the number of all diseases per 1000 residents per year), CO and $SO_2$ are the concentrations of carbon monoxide and sulfur dioxide in the air (in 'threshold limit value' (TLV) units).

The next step of the study involves finding a regression model with two predictor variables that has the maximum value of the determination coefficient $R^2$ among all two-factor models. This model is given by

$$Y = 634 + 1047 \cdot CO - 278 \cdot SO_2 \tag{26}$$

Both predictors in (26) are statistically significant: for the predictor CO, the p-value is equal to 0.00073; for the predictor $SO_2$ we have p=0.021. Equation (26) includes the predictor CO (as expected because this predictor has the maximum coefficient of correlation with Y) and the predictor $SO_2$ (which is unexpected because its coefficient of correlation with Y is very small). The combination of these predictors gives a maximum determination coefficient ($R^2$=0.527) among the all regression models with two predictors. This $R^2$ exceeds considerably the square of the Pearson coefficient of correlation between Y and CO from table 1. It may be stated that $R^2 = 0.527$ is enough high from the epidemiological point of view considering that morbidity is determined by various factors (not only by environmental pollution).

In accordance with the generally accepted understanding (Draper and Smith 1998); Bonita et al. 2006), model (26) could be interpreted as follows: as the concentration of carbon monoxide increases by a unit TLV (threshold limit value), morbidity *increases* by 1047 cases per 1000 residents per year (irrespective of $SO_2$ values), while an increase in the concentration of sulfur dioxide by a unit TLV provides a *decrease* in morbidity by 278 cases (irrespective of CO values).

This interpretation, however, contradicts the epidemiological expectations concerning the effect of pollutants on health: morbidity should increase as pollution increases, and the coefficient for $SO_2$ in equation (26) should be negative. The negative regression coefficient for $SO_2$ in equation (26) does not agree also with the results of a single-factor analysis: in the simple regression equations (24) and (25), both regression coefficients for CO and $SO_2$ are positive.

The cause of the discrepancy between the simple and multiple regression models and the reason for the emergence of model coefficients whose signs contradict the epidemiological expectations is well known: it is the correlation between the pollutants (Draper and Smith 1998). Ehrenberg (Ehrenberg, 1977, section 15.1) also showed that where a strong correlation is present between predictors $x_1$ and $x_2$, one can construct several regression models of similar quality (in terms of the magnitude of $R^2$) whose coefficients differ even in signs (in Ehrenberg's example, the coefficient of correlation between $x_1$ and $x_2$ is equal to 0.790; in our example, $r(CO, SO_2) = 0.729$).

In order to understand how and why regression coefficients change in passing from a simple regression model to multiple regression (in passing from relationships (24)-(25) to (26)), let us use relationships (8) and (9) appearing in the main body of the paper. Here $x_1$ is the toxicant CO, and $x_2$ is $SO_2$. For models (24) – (26), we have: $a_1 = 579$ and $a_2 = 52.5$; $b_1 = 1047$; $b_2 = -278$; $c_{12} = 0.316$ ; $c_{21} = 1.683$; $r(x_1, x_2) = 0.729$.

From the above numerical values it follows that the coefficient $a_1$ in equation (24) is greater than the product $a_2 c_{21}$. Thus, according to (8), the coefficient $b_1$ in (26) is positive while the high value of the correlation coefficient $r^2(x_1, x_2)$ leads to the inequality $b_1 > a_1$. Similar reasoning shows why the coefficient $b_2$ in (26) is negative: The product $a_1 c_{12}$ proves to be greater than $a_2$. It should be emphasized once again: if $r(x_1, x_2)$ were equal to zero (the predictors $x_1$ and $x_2$ were not correlated), then $b_1 = a_1$ and $b_2 = a_2$.

Thus, the cause of a sharp difference between the results of one- and two-factor analyses (24)-(26) is a strong correlation between the concentrations of CO and $SO_2$. Since in large cities a considerable portion of the pollutants is emitted into the atmosphere by automobiles (i.e. there is one source of emissions for several toxicants), the presence of a significant correlation between pollutants is practically unavoidable. It is therefore necessary to find ways to interpret regression models in a meaningful manner when we deal with correlated predictors. One of such ways, as described above, is the inclusion of cross terms into the model.

Consider this method for model (26). If we add a cross term to this model (the product of the concentrations of CO and $SO_2$), we obtain the following equation

$$Y = 204 + 1674 \cdot CO + 36 \cdot SO_2 - 413 \cdot (CO \cdot SO_2).$$  (27)

In model (27), statistically significant is the predictor CO (p=0.00033); for the cross term p=0.115; the linear term $SO_2$ is statistically insignificant (p=0.87). The determination coefficient of model

(27) is equal to $R^2=0.601$, while in the linear model (26) with the same predictors CO and $SO_2$ , it is equal to $R^2=0.527$.

An epidemiological interpretation of model (27) is as follows. Both regression coefficients for the predictor-linear terms are *positive* (morbidity grows with increasing the concentrations of both CO and $SO_2$, which is in agreement with the epidemiological expectations). It is to be noted that the effect produced by the joint action of the two pollutants is a sum of two linear terms (a sum of one-factor effects) and the cross term with a negative sign. It can be said that the effect of the joint action of CO and $SO_2$ is "less than additive". Since the predictor $SO_2$ is statistically insignificant, morbidity Y is determined by a strong negative influence of CO (an increase in CO concentration leads to a sharp increase in morbidity), while the presence of a cross term attenuates the influence of CO (as the concentration of $SO_2$ increases the influence of CO goes down).

As a consequence of correlation between the predictors CO and $SO_2$, only part of the "joint" space of these predictors is experimentally "observable" (the inner space of the ellipse in Fig. 1), while the other part is unobservable (owing to the strong correlation between the toxicant, the unobservable area is rather large). Model (27) is *interpolational* for the observable region of changes in CO and $SO_2$ and *extrapolational* for the unobservable one. For the interpolation region, model (27) describes well the experimental values of morbidity Y (sufficiently high $R^2$); for the extrapolation region, the reliability of model (27) may be found to be low.

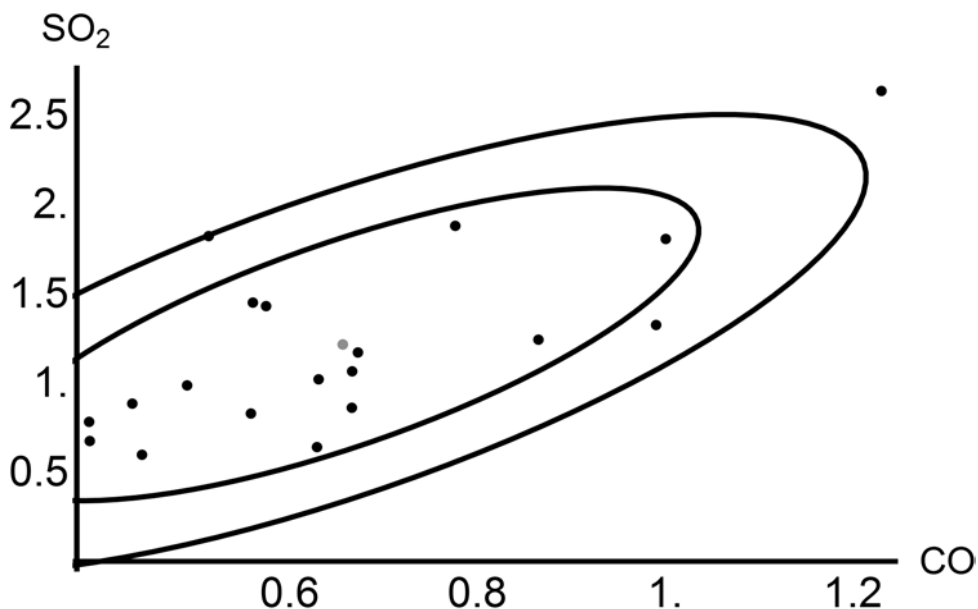

Fig. 1. The scatterplot of the toxicants CO and $SO_2$ for model (27) together with 75% and 95% confidence ellipses (CO and SO2 in TLV units).

Out of the single-factor $Y = Y(x_i \mid \{x_j^0\})$ functions associating the morbidity Y with each toxicant separately, it makes sense to consider the function for one predictor CO since the second predictor, $SO_2$, is statistically insignificant in model (27). The single-factor function for the predictor CO derived from (27) is given by ($SO_2^0$ designates the observable values of the factor $SO_2$)

$$Y(CO \mid SO_2^0) = (204 + 36 \cdot SO_2^0) + (1674 - 413 \cdot SO_2^0)CO \qquad (28)$$

The experimental values of $SO_2^0$ observed in the air of Saint-Petersburg vary within the range of 0.598 to 2.63 in TLV units. Consequently, for any values of $SO_2^0$ the relationship (28) predicts an increase in the morbidity with an increase in the concentration of CO in the form $Y = a_0 + a_1 \cdot CO$ (coefficient $a_1$ is positive). In so doing, an increase in the values of $SO_2^0$ from minimal to maximal leads to a reduction in the coefficient $a_1$ from 1427 to 588 (a substantial decrease in the influence of the toxicant CO on the morbidity as the concentration of $SO_2$ increases). The coefficient $a_0 = 204 + 36 \cdot SO_2^0$, on the contrary, increases with an increase in the concentration of $SO_2$ from 226 to 299.

Thus, allowing for a cross term which renders model (27) more adequate to the experimental data (the inclusion of the cross term increases the model's determination coefficient $R^2$ significantly) also changes the substantive interpretation of the model: instead of the oppositely directed actions of the toxicants in model (26), in which the toxic agents have opposite signs, in (27) we see a significant effect on morbidity from only one toxicant, CO, whose action is attenuated by the cross term containing the toxicant $SO_2$.

If we used, as previously, the method of all regressions including all 12 toxicants and restricting ourselves to two predictors, the second largest-$R^2$ (after model (26)) regression model has the form:

$$Y = 481 + 512 \cdot CO + 278 \cdot Pb. \qquad (29)$$

In model (29), the coefficient of correlation between the predictors $r(CO, Pb) = 0{,}302$ is considerably lower than for predictors in model (26). The signs of both coefficients in (29) are positive and the same as those of the coefficients in simple regression models. Statistically significant in (29) is only the predictor CO (p=0.025). The determination coefficient $R^2 = 0{,}379$ in model (29) is much lower than in model (26).

By incorporating a cross term into model (29) we obtain the model

$$Y = -955 + 2879 \cdot CO + 2367 \cdot Pb - 3382 \cdot (CO \cdot Pb). \qquad (30)$$

In model (30), all predictors are statistically highly significant (p<0.006), and the determination coefficient $R^2$=0.636 is higher than in model (27). Both regression coefficients for predictor-linear terms are positive (morbidity grows as the concentrations of both CO and Pb increase), while the cross term is negative. Thus, the effect of the joint action of CO and Pb is "less than additive" as well as that of *CO* and *SO₂*. Morbidity Y is determined by a strong negative influence of CO and Pb (an increase in CO and Pb concentrations leads to a sharp increase in morbidity), while the presence of the cross term attenuates their combined action.

**Example 2.** A regression model with two uncorrelated predictors.

In this example, we show that even in cases of uncorrelated predictors the interpretation of a regression model is not always based on coefficients $b_i$ for a regression model linear in predictors. Often in order to obtain an adequate model it is necessary to include a cross term, which can play an important part in the description of experimental data and their interpretation.

In the experiment (Varaksin et al. 2014), data were obtained on the effects of lead (Pb) and cadmium (Cd) on a number of indices in laboratory animals. In this example, we analyze the index "Succinate dehydrogenase activity in lymphocytes" (SDH). The experimental data for this index (Varaksin et al. 2014) are shown in Fig. 2.

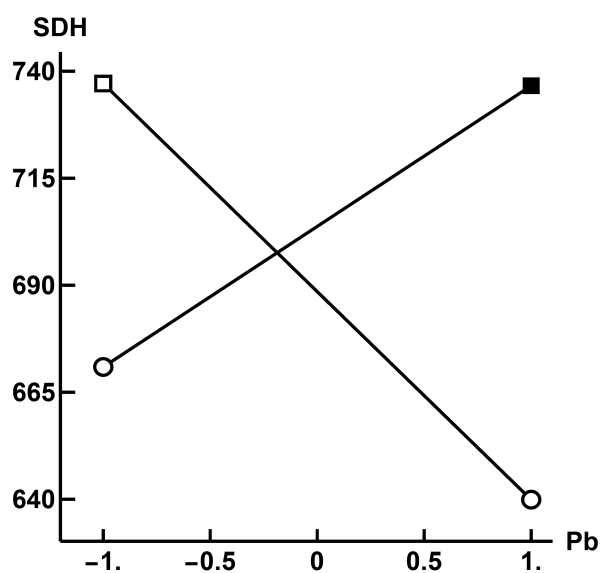

Fig. 2. Experimental mean values of SDH in the control group – empty square (absence of both Pb and Cd); groups with Pb or Cd – empty circle (only one toxicant is present); group with presence both Pb and Cd – black square.

For the SDH index we compare a model linear in predictors of type (5) and a model with a cross terms of type (13). Unlike example 1, in this case we deal with a designed experiment in which the toxicants Pb and Cd act independently of each other so that the corresponding predictor variables are not correlated.

In accordance with the theory of design of experiment (Cox and Reid 2000, Hinkelmann and Kempthorne 2008), doses of toxicants are designated with special codes: The minimum value of a toxicant dose is coded as «–1», and the maximum value as «+1». In the experiment under consideration (Varaksin et al. 2014), the minimum dose of the toxicants was equal to zero (code = –1), and the maximum one to $0.05LD_{50}$ (code = +1). In these codes, the regression model linear in predictors is given by:

$$Y(=SDH) = 693.0 - 4.70\ Pb + 4.49\ Cd. \qquad (31)$$

From (31) it follows that lead *reduces* SDH while cadmium *increases* it, which is not in agreement with the experimental data (in the experiment, both toxic agents reduced SDH relative to the control group - Fig. 2). The coefficient of determination in model (31) is extremely small: $R^2=0.035$. The plot of SDH against the dose of lead for various doses of cadmium according to model (31) is shown in Fig. 3a. We can see essential differences between model (31) and the experiment.

Let us add a cross term to model (31) as follows (Varaksin et al. 2014):

$$Y(=SDH) = 693.0 - 4.70\ Pb + 4.49\ Cd + 43.92\ (Pb*Cd). \qquad (32)$$

The terms linear in predictors in the equation with a cross term (32) have the same coefficients as in equation (31). This is an obvious consequence of the fact that the predictors Pb and Cd are not correlated in this experiment. The determination coefficient $R^2=0.445$ in model (32) is substantially higher than in model (31).

Fig. 3*b* demonstrates explicitly that the inclusion of a cross term changes radically the form of the SDH-'lead dose' dependence for various doses of cadmium. Note that the results of the calculations by model (32) are in very good agreement with the experimental data (the estimates are virtually the same as the mean experimental values of SDH in different groups).

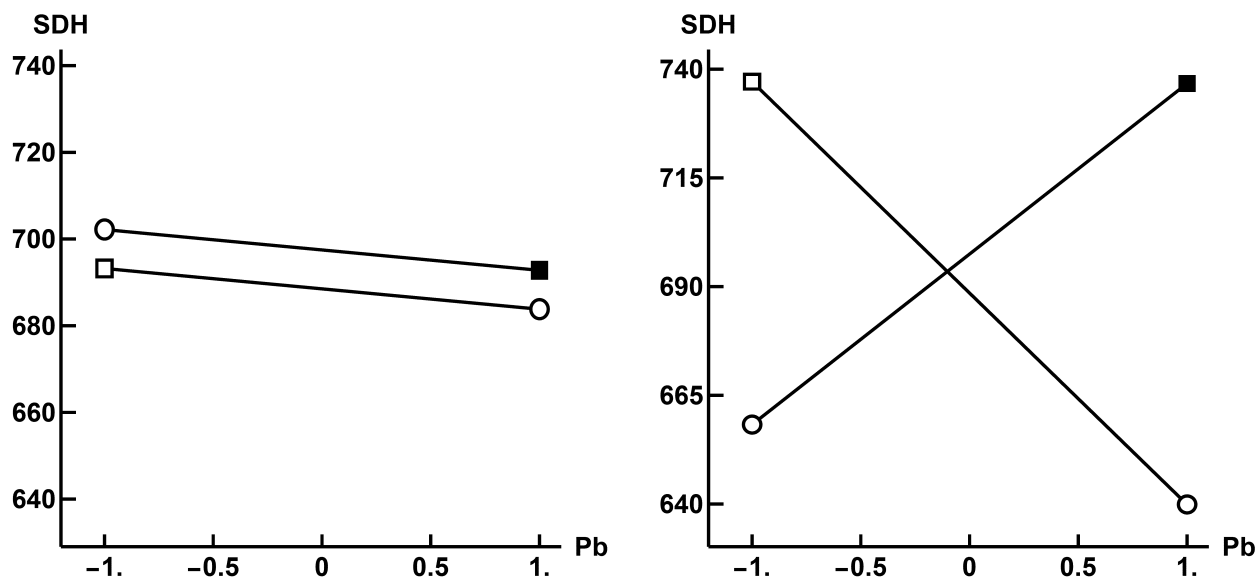

Fig. 3. The dependence of SDH activity on lead dose for various doses of cadmium. a) predicted results according to linear model (31); *b*) model (32) with a cross term. The designations are the same as in Fig. 2.

Consider model (32) interpretation. The conditional one-factor functions describing the relationships between SDH and doses of lead and cadmium are given by:

$$Y(Pb\,|\,Cd^0) = (693,0 + 4,49\,Cd^0) + (43,92\,Cd^0 - 4,70)Pb\,, \qquad (33)$$

$$Y(Cd\,|\,Pb^0) = (693,0 - 4,70\,Pb^0) + (43,92\,Pb^0 + 4,49)\,Cd\,. \qquad (34)$$

From (33) it follows that in the absence of cadmium ($Cd^0 = -1$), increasing the dose of lead reduces SDH, which is consistent with the experiment. On the contrary, in the presence of cadmium ($Cd^0 = +1$), increasing the dose of lead increases SDH, which is also consistent with the experiment. Relationship (34) is interpreted similarly. Thus, SDH has the maximum value in the absence of both toxicant agents. The corresponding code values of the agents are Pb= –1 and Cd= –1, while the estimate of the response SDH = 737.1 (it matches the mean experimental value). For an isolated action of lead when the dose of cadmium is equal to zero (codes Pb= +1 and Cd= –1), the response is SDH = 639.9 (i.e. the presence of lead reduces SDH). For an isolated action of cadmium when the dose of lead is equal to zero (codes Pb= –1 and Cd= +1), the response is SDH = 658.3 (cadmium, too, reduces SDH). Finally, for a joint action of lead and cadmium (codes Pb= +1 and Cd= +1), SDH =736.7 (the mean experimental value of SDH = 736.6). Thus, the two toxicants, each of which reduces SDH when acting alone, restore the level of SDH-activity when acting together practically to the level in the control group. Such behavior of the SDH-activity index may be interpreted as antagonism (one toxicant mitigates the action of the other). SDH was observed to behave exactly like this in the experiment, and it can be described by a regression model that always allows for a cross term.

Fig. 4 shows a response surface described by model (32). The sectioning of this surface with the planes $Cd = Cd^0$ or $Pb = Pb^0$ gives lines described by formulas (33) and (34). The response surface (Fig. 4) has a rather complex form but, nevertheless, the single-factor functions are linear functions of the Pb and Cd concentrations. Because the predictors Pb and Cd are not correlated, the entire region of dose combinations (Pb; Cd) in Fig. 4 is interpolational (compare with Fig. 1, in which a considerable part of the plane CO - SO$_2$ does not contain any experimental points). It is to be emphasized once again that for interpreting regression models it is essential to analyze the regression function (response surface) as a whole rather than separate coefficients of the regression model.

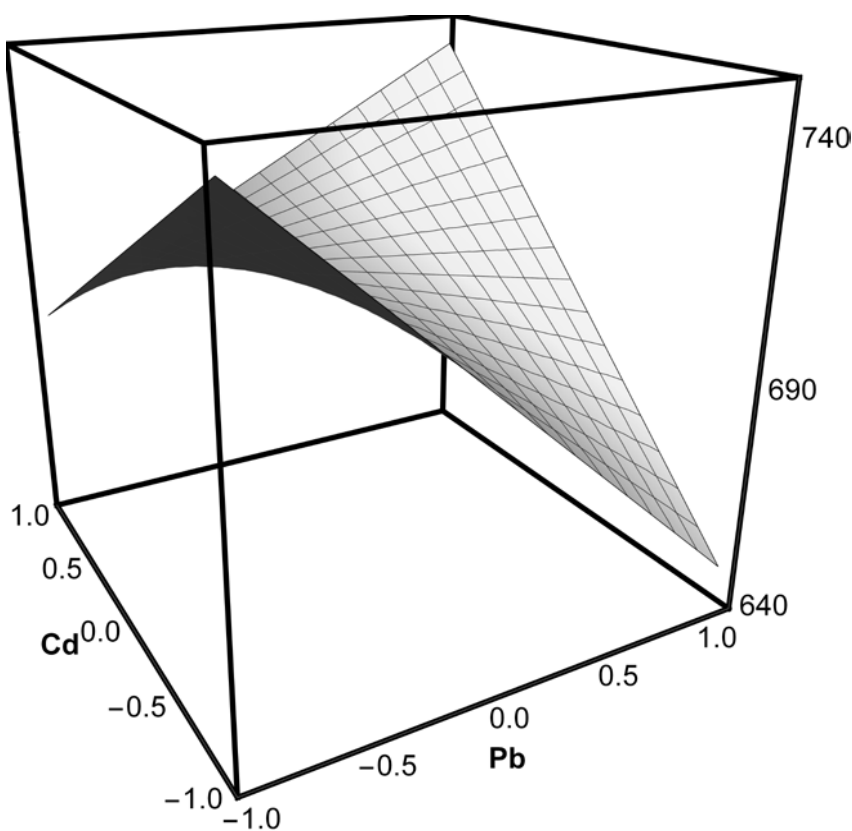

**Fig. 4.** The response surface for SDH – equation (32). The vertical axis shows SDH values.

**Example 3.** A multifactorial regression model and interpretation of the joint action of predictors.

In the examples considered above, the number of predictors is two (three, allowing for the cross term). Let us consider an example of a regression model containing more than two predictors and its interpretation. To this end, it is necessary to have a dataset with a greater number of observations than in examples 1 and 2. In this example, we consider regression models relating the prevalence of tuberculosis (TB) in 53 cities and towns of Sverdlovsk Region (Middle Urals, Russia) to TB risk factors.

The TB risk factors at hand (Socio-economic situation…, 2004; Varaksin and Zaikina, 2010), can be divided into three groups:

1) housing quality (proportion of families living in accommodation with centralized cold and hot water supply and heating);

2) socio-economic indicators: income per family member (thousands of rubles a month), number of clinicians (clinicians of all specialisms per 1000 residents), proportion of top-qualification clinicians (% of all clinicians), apartment size (square meters per family member), unemployment (as percentage of working-age population).

3) environmental pollution indicators (proportion of drinking water samples failing to meet the chemical and micro-biological sanitary standards).

Note that the signs of all predictors in the simple regression models correspond to the epidemiological expectations: the prevalence of tuberculosis decreases with improvements in housing quality, income and access to qualified health care, and with an increase in apartment size; the prevalence of tuberculosis increases with growth in unemployment and worsening of drinking water quality.

Let us construct a multiple linear regression model in accordance with the recommendations described in section 5 of the paper. First we need to find, among all multiple regression models with four predictors, their squares and cross terms (totally 14 variables), a model with the maximum $R^2$. It is a model with the following variables:

$x_1$ – proportion of top-qualification clinicians (as percentage of total number of clinicians);

$x_2$ – proportion of apartments equipped with centralized water supply (hereinafter «water supply» for brevity);

$x_3$ – apartment size (square meters per tenant);

$x_4$ – income (thousands rubles per resident per month).

Table presents the range of experimental values of the main factors $x_1$ - $x_4$.

Table. Experimental ranges of variation of Y and X.

| Factor | | Minimum | 25 Quartile | Mean | 75 Quartile | Maximum |
|---|---|---|---|---|---|---|
| Tuberculosis, number of patients per 100,000 population | Y | 34.3 | 79.4 | 101.0 | 113.3 | 220.0 |
| Proportion of top-qualification clinicians, % | $x_1$ | 39.3 | 57.8 | 63.8 | 69.3 | 80.7 |
| Water supply, % | $x_2$ | 16.2 | 54.0 | 66.3 | 82.4 | 98.7 |
| Apartment size, $m^2$ per tenant | $x_3$ | 16.7 | 18.5 | 19.9 | 21.2 | 24.4 |
| Income, thousands of rubles a month | $x_4$ | 2.15 | 4.42 | 6.07 | 7.47 | 11.4 |

As a consequence of containing a large number of variables (totally 14 terms), the model has few statistically significant terms: these are the cross term $(x_1 \cdot x_2)$ with $p < 0.05$ and the linear term $x_1$ with $p < 0.10$. The determination coefficient of the model is rather high: $R^2 = 0.602$, and the $R^2$, adjusted to degrees of freedom is $R^2_{adj} = 0.456$.

At the second step of model construction, the number of predictors is reduced using the backwards stepwise regression procedure. What remains is all four linear terms and two cross terms: $(x_1 \cdot x_2)$ and $(x_1 \cdot x_3)$. The result of the second step is:

$$Y = 1175 - 15.08 \cdot x_1 - 3.411 \cdot x_2 - 39.62 \cdot x_3 - 4.810 \cdot x_4 + 0.0504 \cdot (x_1 \cdot x_2) + 0.565 \cdot (x_1 \cdot x_3) \qquad (35)$$

In this model with six factors (four linear and two cross terms), all predictors are statistically significant. The determination coefficient $R^2 = 0.542$ is a little lower than previously; however, the adjusted $R^2_{adj} = 0.483$ is higher.

Let us see how the TB prevalence changes with changes in the risk factors. In model (35), the most statistically significant factor is «Proportion of top-qualification clinicians» (hereinafter «proportion of clinicians» for brevity). It also appears in both of the significant cross terms.

According to the multiple regression model (35), TB prevalence as a function of proportion of clinicians (factor $x_1$) is given by

$$Y = T_0 + T_1 x_1 \ , \qquad (36)$$

where the coefficients for regression (36) are functions of the concomitant risk factors $x_2$, $x_3$ and $x_4$:

$$T_0 = 1175 \ - \ 3.411 \, x_2^o - 39.62 \, x_3^o - 4.810 \ x_4^o, \qquad (37)$$

$$T_1 = - \ 15.08 + 0.0504 \, x_2^o + 0.565 \ x_3^o, \qquad (38)$$

and $x_2^o, x_3^o, x_4^o$ are the fixed values of the factors for which the coefficients $T_0$ and $T_1$ are calculated in model (36).

Using formulas (36)-(38), we can compute the dependence of prevalence on the proportion of clinicians for any combination of the concomitant factors (using the values of $x_1^o, x_2^o, x_3^o, x_4^o$ from Table). Thus, if all factors $x_2^o, x_3^o, x_4^o$ are at the level of the 25$^{th}$ quartile, then the coefficients $T_0$ (37) and $T_1$ (38) are: $T_0 = 237.6$ and $T_1 = -1.921$ (value of the coefficient $T_1$ is close to that in simple regression, obtained using one predictor $x_1$: $a_1 = -1.889$). If the factors $x_2^o, x_3^o, x_4^o$ are at the medium level, then $T_0 = 132.1$ and $T_1 = -0.510$. We can see that as the values of the factors $x_2^o, x_3^o, x_4^o$ increase, the strength of association between TB prevalence and proportion of clinicians decreases (a decrease modulo coefficient $T_1$). This state of things is explainable: an increase in the values of $x_2^o, x_3^o, x_4^o$ means an improvement in living conditions with the result that the effect of one of these risk factors (proportion of clinicians) on the development of tuberculosis reduces. Finally, if the factors $x_2^o, x_3^o, x_4^o$ are at the level of the 75$^{th}$ quartile, then $T_0 = 14.2$ and the coefficient $T_1$ becomes positive $T_1 = +1.11$. This may seem unexplainable (the prevalence of tuberculosis should not grow with the proportion of clinicians). To explain this phenomenon, consider the coefficients of correlation between the factors $x_2, x_3$ that appear in formula (18) for the coefficient $T_1$. This coefficient is small: $r(x_2, x_3) = 0.129$; hence, increasing $x_2$ does not actually increase the factor $x_3$ proportionally. Similarly, increasing in $x_3$ does not strictly follow the increasing of $x_2$. We calculated the coefficient $T_1$ (18) for $x_2 = 75$th quartile and took for $x_3$ the mean value derived from the dataset for the values of $x_2$ appearing in the 75th quartile. The coefficient $T_1$ then becomes equal to zero. If we now take the value of $x_3$ equal to the 75th quartile and $x_2$ to the mean value derived from the dataset for this sample, we obtain $T_1 = -0.012$.

Thus, correlation (or absence of any correlation) between predictors has an explicit effect on the epidemiological interpretation of a regression model and thus should be taken into account when defining the region of predictor values that is going to be used for data analysis and predicting.

It is possible to derive from the multiple-factor model (35) a single-factor dependence of TB prevalence depending on other predictors, for instance, on the availability of centralized water supply (factor $x_2$). Now, instead of relationships (36)-(38) we have

$$Y = T_0 + T_2 x_2 \ , \tag{39}$$

where

$$T_0 = 1175 - 15.08 x_1^o - 39.62 x_3^o - 4.810 x_4^o + 0.565 x_1^o x_3^o \ , \tag{40}$$

$$T_2 = -3.411 + 0.0504 x_1^o \ , \tag{41}$$

$x_1^o, x_3^o, x_4^o$ are the fixed values of the factors for which the coefficients $T_0$ and $T_2$ in equations (39)-(41) are calculated. Unlike the coefficient $T_1$ from (18), which is dependent on two concomitant factors $x_2$ and $x_3$, the coefficient $T_2$ in (41) depends on one concomitant factor $x_1$. From (41) it follows that the coefficient $T_2$ decreases (in modulus) as the proportion of clinicians $x_1^o$ increases. Thus, for the values of $x_1^o$ equal to the 25th quartile, the mean value and the 75th quartile ($x_1^o = 57.8$; 63.8; 69.3), respectively, we have $T_2 = -0.498$; $T_2 = -0.195$ and $T_2 = +0.0817$, respectively. The situation is similar to the one considered above: as the living conditions improve according to some indicators (in this case, an increase in the proportion of clinicians), the effects of other indicators (in this case, availability of centralized water supply) on TB prevalence decrease.

Note one more fact. The above-calculated values of the coefficient, $T_2 = -0.498$; $T_2 = -0.195$ and $T_2 = +0.0817$, become less (in modulus) than the coefficient $a_2$ in the simple regression equation $Y = a_0 + a_2 x_2$ associating TB prevalence to the only factor $x_2$: $a_2 = -0.858$. Thus, allowing for the concomitant variable (factor $x_1$ in this particular case) attenuates the association between TB prevalence (Y) and centralized water supply (factor $x_2$).

A similar interpretation follows for the association between TB prevalence and factor $x_3$ (apartment size); as well as the association for $x_2$, this one is also dependent on one concomitant factor $x_1$. At the same time, the relationship between TB and factor $x_4$ (income) does not at all depend on concomitant factors within the statistical significance limits and is determined by the relationship $Y = T_0 + T_4 x_4$ , where $T_0$ depends on $x_1$, $x_2$, $x_3$ and $T_4 = -4.810$ (for comparison $a_4 = -8.617$).